\documentclass[%
 reprint,
 amsmath,amssymb,
 aps,
]{revtex4-1}

\usepackage{graphicx}
\usepackage{dcolumn}
\usepackage{bm}
\usepackage{hyperref}
\usepackage[dvipsnames]{xcolor}
\usepackage{stackrel}
\usepackage[inline]{enumitem}

\usepackage{xr}

\usepackage{tikz}
\usetikzlibrary{arrows.meta} 
\usetikzlibrary{calc}
\usepackage{pgfplots}

\usepackage{epstopdf}

\newcommand{\dd}[2]{\frac{d #1}{d #2}}
\newcommand{\e}[1]{^{(#1)}}

\definecolor{myblue}{rgb}{0.2, 0.5, 0.91}  
\definecolor{purpleheart}{rgb}{0.41, 0.21, 0.61} 
\definecolor{mygreen}{rgb}{0.2, 0.7, 0.2} 
\definecolor{carnelian}{rgb}{0.7, 0.11, 0.11} 
\definecolor{amber}{rgb}{1.0, 0.75, 0.0} 
\definecolor{beige}{rgb}{0.96, 0.96, 0.86}
\definecolor{corn}{rgb}{0.98, 0.93, 0.36}
\definecolor{darktangerine}{rgb}{1.0, 0.66, 0.07}

\definecolor{arylideyellow}{rgb}{0.91, 0.84, 0.42} 
\definecolor{cosmiclatte}{rgb}{1.0, 0.97, 0.91}

\definecolor{almond}{rgb}{0.94, 0.87, 0.8}  
\definecolor{antiquewhite}{rgb}{0.98, 0.92, 0.84}
\definecolor{floralwhite}{rgb}{1.0, 0.98, 0.94}

\newcommand{\csquare}{cosmiclatte} 
\newcommand{\ca}{myblue} 
\newcommand{\cb}{purpleheart} 
\newcommand{\cc}{mygreen}

\tikzset{interaction/.style={line width=2pt}
}
\tikzset{arrow/.style={thick,color=black,shorten <=3pt,shorten >=3pt
}
}

\tikzset{
pics/mysquare/.style args={#1}{
 code = {
   \draw[black,fill=\csquare] (0,0) rectangle (1,1) node[midway] {\Large #1}; 
   }
 }
}

\tikzset{
pics/my4square/.style args={#1}{
 code = {
   \pgfmathsetmacro\a{int(#1+1)}
   \pgfmathsetmacro\b{int(#1+2)}
   \pgfmathsetmacro\c{int(#1+3)}
   \pic at (0,0) {mysquare=#1};
   \pic at (1,0) {mysquare=\a};
   \pic at (0,1) {mysquare=\b};
   \pic at (1,1) {mysquare=\c};
   \draw[interaction,\ca] (1,0) -- (1,2);
   \draw[interaction,\ca] (0,1) -- (2,1);
   }
 }
}

\tikzset{
pics/my16square/.style args={#1}{
 code = {
   \pgfmathsetmacro\d{int(#1+4)}
   \pgfmathsetmacro\e{int(#1+8)}
   \pgfmathsetmacro\f{int(#1+12)}
   \pic at (0,0) {my4square=#1};
   \pic at (2,0) {my4square=\d};
   \pic at (0,2) {my4square=\e};
   \pic at (2,2) {my4square=\f};
   \draw[interaction,\cb] (2,0) -- (2,4);
   \draw[interaction,\cb] (0,2) -- (4,2);
   }
 }
}

\tikzset{
pics/my64square/.style args={#1}{
 code = {
   \pgfmathsetmacro\g{int(#1+16)}
   \pgfmathsetmacro\h{int(#1+32)}
   \pgfmathsetmacro\i{int(#1+48)}
   \pic at (0,0) {my16square=#1};
   \pic at (4,0) {my16square=\g};
   \pic at (0,4) {my16square=\h};
   \pic at (4,4) {my16square=\i};
   \draw[interaction,\cc] (4,0) -- (4,8);
   \draw[interaction,\cc] (0,4) -- (8,4);
   }
 }
}

\tikzset{
pics/square2/.style args={#1/#2}{
 code = {
   \draw[black,fill=#2] (0,0) rectangle (1,1) node[midway] {\Large #1}; 
   }
 }
}

\pgfplotsset{compat=1.18}

\begin{document}

\preprint{APS/123-QED}

\title{Hierarchical self-assembly for high-yield addressable complexity at fixed conditions }

\author{Miranda Holmes-Cerfon}
 \email{holmescerfon@math.ubc.ca}
\affiliation{%
Department of Mathematics, University of British Columbia\\
1984 Mathematics Rd, Vancouver, BC, V6T 1Z1, Canada
}%

\author{Matthieu Wyart}%
 \email{matthieu.wyart@epfl.ch}
\affiliation{
Department of Physics and Astronomy\\
Johns Hopkins University, Baltimore, MD, USA\\
Institute of Physics\\
\'Ecole Polytechnique Fédérale de Lausanne, Lausanne, CH-1015, Switzerland
}

\date{\today}

\begin{abstract}
There is evidence that the self-assembly of complex molecular systems often proceeds hierarchically, by first building subunits that later assemble in larger entities, in a process that can repeat multiple times. Yet, our understanding of this phenomenon and its performance is limited. Here we introduce a simple model for hierarchical addressable self-assembly, where interactions between particles can be optimised to maximise the fraction of a  well-formed target structure, or yield.  We find that  a hierarchical strategy leads to an impressive yield up to at least five generations of the hierarchy, and does not require a cycle of temperatures as used in previous methods. 
High yield is obtained when the microscopic interaction decreases with the scale of units considered, such that the total interaction between intermediate structures remains identical at all scales.  
We provide thermodynamic and dynamical arguments constraining the interaction strengths where this strategy is effective. Overall, our work characterizes an alternative strategy for addressable self-assembly at a fixed temperature, and provides insight into the mechanisms sustaining hierarchical assembly in biological systems. 
\end{abstract}

\maketitle

\section{Introduction}

The ribosome, the  central machinery that translates messenger RNA into proteins, illustrates the surprising ability of multi-components systems to self-assemble. It is a protein complex  made of  $\sim100$ proteins (in addition to  ribosomal RNA), containing around $10^4$ amino acids. Considering the astronomical number of possible configurations these amino acids could take, how does this system self-assemble correctly? Clearly, this process is simplified by the fact that amino-acids first fold into proteins, which later form a complex. There is evidence that such a hierarchical strategy, where small units first form before assembling into  larger ones, is at play in the folding of individual proteins itself \cite{dill2008protein}. In this view, proteins  first nucleate secondary structures such as beta sheets or alpha helices \cite{baldwin1999protein} or more complex ``foldons'' \cite{fersht1997nucleation} which later organize into the full tertiary structure.  Hierarchical self-assembly is also used in chemistry to obtain supra-molecular structures, typically held together with hydrogen bonds or Van der Waals interactions, much weaker than the covalent bonds organizing molecules \cite{elemans2003mastering}. More generally,  a hierarchical organization is central to a variety of complex systems and processes, from engineering to the composition of companies from smaller entities \cite{simon1962architecture}. Despite its importance and ubiquity, understanding the principles governing hierarchical assembly remains a challenge.  

Such principles could be used and tested in the context of ``addressable'' self-assembly, where each unit of a target structure is distinct and must assemble into a particular location \cite{Rothemund.2012,Rothemund.2006,Sacanna.2013,Wang.2012,Rogers.2015,Rogers.2016,Hong.2017,murugan2015multifarious}. 
Using DNA strands as building blocks with  highly specific interactions, experimentalists have so realized highly complex structures containing up to tens of thousands of units \cite{Wei.2012,Ke.2012,Ong.2017,evans2024pattern}. 
A challenge is to 
find the particular building blocks \cite{king2024programming} or protocols \cite{goodrich2021designing,bupathy2022temperature,engel2023optimal} such that the system assembles into a desired structure. 
Experiments operate in the regime where all pair interactions have roughly the same strength \cite{Hormoz.2011}, although the importance of displaying a range of interactions has been proposed in various contexts \cite{jhaveri2024discovering,kaplan2014building}. When these interactions are strong, or equivalently the temperature is low, the system assembles many partially-formed fragments of the target structures, in a phenomenon known as ``monomer starvation''. 
These fragments must break apart to assemble into copies of the target structure, but they cannot do so on accessible timescales -- although the target structure has the lowest free energy, it is not kinetically accessible. In contrast, with weaker interactions  
the target state 
no longer has the lowest free energy. %
One solution, identified experimentally and later rationalized theoretically \cite{Reinhardt.2014,Jacobs.2015} is to anneal temperature: start at a high temperature where nucleation is rare, then slowly lower temperature to grow the nuclei, repeating until one forms a target structure. This nucleation strategy can be enhanced if some parts of the assembly present a larger chemical potential \cite{murugan2015undesired}.
While this protocol has worked 
well as a method to form one or a small number of copies of a target structure,  
it wastes monomers, since many monomers do not assemble into a target structure, and furthermore it requires a precise temperature protocol. 

We wish to understand whether hierarchy can be used to achieve high-yield assembly at fixed conditions, without annealing. Most previous studies of hierarchical assembly required some kind of staged assembly, where the experimental conditions or protocols change with time \cite{Park.20069d9,Grunwald.2014,Whitelam.2015,Pfeifer.2016,Wagenbauer.2017,Zhang.2017,mcmullen2022self}, which is  not always possible in biological contexts. A handful of recent studies have developed experimental  systems that assemble via two stages of hierarchy at equilibrium \cite{freeman2018reversible,hayes2021encoding,Zou.2023,Jiang.2024}. This is often achieved using a mix of  ``strong'' bonds, and ``weak'' bonds which become stronger due to geometric organization once the strong bonds have formed \cite{Gartner.2024}.

Here we introduce a minimal model of assembly at equilibrium, which we study up to five hierarchical steps. Our central results are that: (i) hierarchical assembly is possible in equilibrium, with fixed conditions; (ii) efficient assembly requires the scale of interaction to decrease with size, so that the binding energy of meta-particles is approximately independent of scale, (iii) the dynamics of self-assembly is itself hierarchical,  the characteristic time  at which meta-particles are formed scales with their length  and (iv) we derive some constraints on the interaction strength for hierarchical self-assembly to properly occur, summarizing our results in a phase diagram.

\begin{figure*}
\centering
\includegraphics[width=\linewidth]{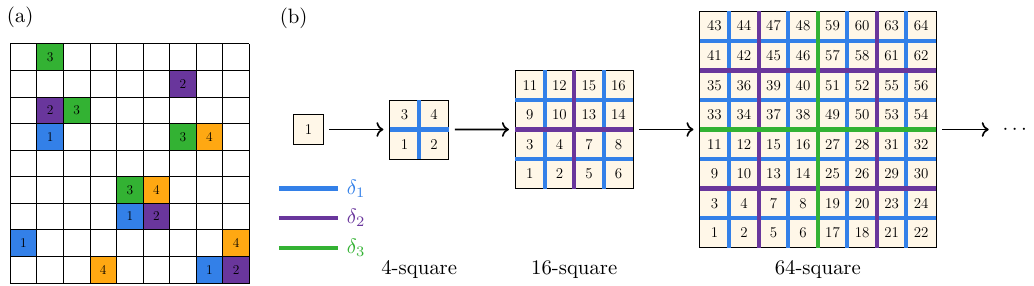}
\caption{Schematic of our model and the hierarchical assembly scheme. (a) Our model considers particles on a square lattice interacting with pairwise energies. Clusters can move as a unit and sometimes break apart. 
The example shown contains enough monomers to make $N_c{=}4$ copies of a 4-square, but only one 4-square has fully formed.  (b) Building large squares with addressable complexity using hierarchical interactions. Square monomers with sticky sides assemble into 4-squares with strong edge interactions $\delta_1$, 4-squares assemble into 16-squares with weaker individual edge interactions $\delta_2$, 16-squares assemble into 64-squares using even weaker edge interactions $\delta_3$, and so on. We expect good assembly when the interactions are chosen hierarchically, as $\delta_2{=}\delta_1 / 2$, $\delta_3{=}\delta_2/2$, $\delta_4{=}\delta_3/2$, etc, for suitable values of the strongest energy $\delta_1$. This way, the energy for gluing together $4^k$-squares into $4^{k+1}$-squares ($\delta_1$), is the same as for gluing together monomers into 4-squares.
}\label{fig:hier}
\end{figure*}

\section{Hierarchical assembly with sticky squares}

\subsection{Model}

We consider a system of sticky squares (monomers)  on a two-dimensional square lattice with periodic boundary conditions, as depicted in Fig. \ref{fig:hier} (left). 
  We assume that monomer $i$ interacts with monomer $j$ with an interaction energy $0$ if the squares are not touching, $\infty$ if the squares overlap, and $-\delta_{ij}^{\vec{l}}$ if the squares are exactly touching with separation $\vec{l} \in \{(1,0), (-1,0), (0,1), (0,-1)\}$. 
  The dependence on $\vec{l}$ models directional interactions.
The target structure is an \emph{$n$-square} containing $n=2^k\times 2^k=4^k$ monomers (Figure \ref{fig:hier}) where $k$ is some integer, with each monomer in a distinct location. 
Our goal is to make $N_c$ copies of this target structure at a constant temperature $T$. We measure interaction energies in units of $k_BT$ so the temperature will no longer enter our discussion. We  simulate a collection of $nN_c$ monomers with interactions chosen such that $\delta_{ij}^{\vec{l}} > 0$ if $i$ and $j$ are copies of neighbouring monomers with separation $\vec{l}$ in the target structure. We denote by $\rho$ the concentration of any specified monomer, which is assumed equal for all monomers. All our simulations are run in a box with side length $L$ and volume $V=L^2$, at volume fraction $\phi = nN_c/V \approx 0.05$; note that $\phi = 4^k\rho$. 

 The dynamics is modeled using the Virtual Move Monte Carlo algorithm \cite{Whitelam.2007}, whose equilibrium converges to the Boltzmann distribution, and which preserves many natural dynamical features when time is measured in Monte Carlo MC sweeps (one MC sweep is $nN_c$ Monte Carlo moves). Importantly, this algorithm allows moving clusters as a unit, merging clusters, and breaking apart clusters into sub-clusters.  The average diffusion coefficient of a cluster may be a prescribed function of the cluster's size or shape. We use an approximation to 3d Stokes' drag, which implies a $4^k$-square diffuses approximately with diffusion coefficient $D_k \approx 1/\sqrt{n}=1/2^k$. (We verify in the SI, Section \ref{sec:diffusion}, 
 that other models for diffusion do not alter the success of our proposed scheme.) To speed up the code, we enforce that monomers cannot rotate, which we expect not to change results qualitatively \cite{Evans.2017}. Consistent with the physics of micronscale particles, we treat the particles as ``sticky'' and do not include any barriers to reacting, so monomers interact as soon as they are in contact. We note however that including such reaction barriers gives another parameter for sculpting the kinetics of self-assembly \cite{benoist2025high}. 

 Details of the algorithm and its implementation, including how diffusion coefficients are enforced, are described in  \cite{Whitelam.2007}. Our code is adapted from \cite{VMMC_3} and a version is available on Github at \cite{code}.

Our model bears similarities in its thermodynamic properties to other computational models used to study addressable self-assembly, e.g. \cite{murugan2015multifarious}. It is notably different from most other studies in two aspects. One is its dynamics; many studies (with a handful of exceptions) only allow monomers to diffuse. A second is that we work in the canonical ensemble, with fixed particle number; rather than the grand canonical ensemble. The canonical ensemble is the setting that experiments operate in, and this choice allows us to address one of the key issues facing experimentalists, namely monomer starvation.

We aim to choose $\delta_{ij}^{\vec{l}}$ to make the system assemble hierarchically, and to explore when this makes the overall yield of the target structure high. 
We measure the yield $y_k$ of $4^k$-squares by the total number $N_f$ of  \emph{perfectly formed} target structures, $y_k=N_f/N_c$, measured at a given time. We denote by $\Delta_k \equiv  1-y_k$ the fraction of structures that do not correspond to the target. 

\begin{figure*}
\centering
\includegraphics[width=0.9\linewidth]{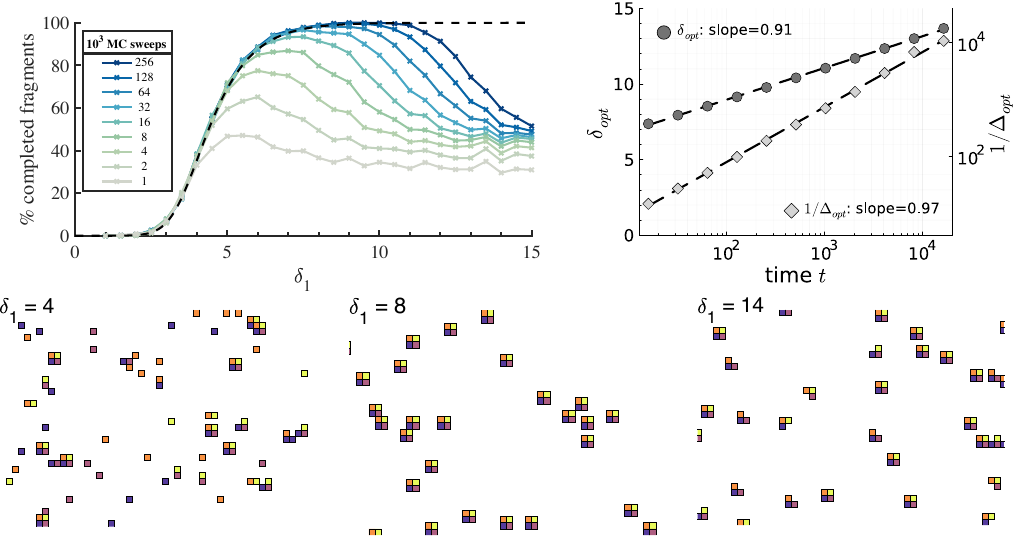}
\caption{Building 4-squares with varying side energies $\delta_1$. Top left: the percentage of fully-completed 4-squares from simulations at different times (markers/solid lines). Black dashed line shows the equilibrium fraction of 4-squares, from Eq.(S4). 
Simulations considered $N_c=100$, $L=89$, and for each $\delta_1$ were repeated 12 times to obtain an average yield.  Bottom: snapshots of individual simulations after 128 MC sweeps, showing about 1/9th of each system.  For small $\delta_1$, the system is in equilibrium and contains a mixture of fragment sizes. For medium $\delta_1$, the system is nearly 100\% 4-squares. For large $\delta_1$, the system forms a mixture of 3-mers and 4-squares. The 3-mers cannot fit together to form 4-squares without breaking apart, which happens on a longer timescale, so the system is kinetically trapped. 
Top right: the optimal interaction energy $\delta_{\rm opt}$ which maximizes $c_4(t)$, the concentration of 4-squares at time $t$, and the corresponding inverse fraction of errors $\Delta^{-1}_{\rm opt}(t) = (1-c_4(t)/\rho)^{-1}$, found by solving nonlinear rate equations (S8) 
at $\phi=0.05$. Dashed lines are the best-fit lines on a log-linear and log-log scale respectively; their slopes are shown in the legends and are in good agreement with the unit slopes predicted by Eqs. \eqref{e2}, \eqref{e3}. 
}\label{fig:n4}
\end{figure*}

\subsection{Assembling 4-squares} 

We first study how  elementary meta-particles, 4-squares,  assemble.  We set the native interaction energies to $\delta_1$ (in units of $k_BT$) and all other interactions to 0. 
Fig. \ref{fig:n4} shows the yield $y_1$ as a function of $\delta_1$ at different times.
At long times, the yield must approach its equilibrium value, the black dashed curve in Fig. \ref{fig:n4} derived below. For a fixed time, the yield increases to a maximum at intermediate values of $\delta_1$, then  decreases. 
The optimal yield increases with time, as does the value of $\delta_1$ which produces it. 

The decrease in yield at long times occurs because for large $\delta_1$, the system forms many fragments of three monomers (3-mers) as illustrated Fig. \ref{fig:n4}. 
The 3-mers must break apart to fit together into 4-squares, but doing so requires a long timescale $\sim e^{\delta_1}$ for large $\delta_1$. The system is kinetically trapped at intermediate times. 

{\it Theory:} An exact formulation of the thermodynamics and kinetics of the formation of 4-squares is presented in the SI, Section \ref{sec:theory}. Here we emphasize key facts. In the regime of high yield, the dominant species at equilibrium is the fully formed 4-squares of concentration $c_4$, followed by individual monomers of concentration $c_1$. Thus the decrease in yield is $\Delta_1\approx c_1/\rho$ while $c_4\approx \rho$. Imposing that the reaction whereby 4 single monomers form a 4-square satisfies detailed balance leads to $c_1=e^{-\delta_1}c_4^{1/4}$, which implies:

\begin{equation}
\label{e1}
    \Delta_1\approx e^{-\delta_1} \rho^{-3/4}.
\end{equation}

Two time scales govern the kinetics of the assembly process: the diffusion time for monomers to meet, $\tau_D\sim (D_0\rho)^{-1}$, and the unbinding time for a bond to break, $\tau_B\sim e^{\delta_1}/D_0$ (recall $D_0$ is the diffusion coefficient of a monomer). When the ratio $f\equiv \tau_B/\tau_D=e^{\delta_1}\rho$ is large, the concentration $c_3$ of triangles remains large (of order $\rho$) during the assembly process, whose duration is of order $\tau_B$. This effect is apparent at the bottom of Fig.\ref{fig:n4}, where an abundance of triangles is visible at the considered time scale. As we shall see below, this situation is detrimental in the hierarchical assembly, as it can lead to the presence of defects. For 4-squares it is not an issue: if $\rho$ is fixed and one is given a large time $t$ to assemble, then the optimal yield is obtained by choosing $\delta_1$  such that $\tau_B\sim t$, corresponding to 
\begin{equation}
\label{e2}\delta_1\approx \ln(D_0 t).
\end{equation}
This leads to an error that decreases rapidly with the assembly time, as $\Delta_1\sim 1/(t\rho^{3/4})$ or equivalently:
\begin{equation}
\label{e3}
\Delta_1^{-1}\approx D_0t\rho^{3/4}.
\end{equation}
To test the time-dependence of these two scaling laws, we compute the value of $\delta_1$ which optimizes the concentration of 4-squares at different times $t$, by solving nonlinear rate equations for the different structures entering the assembly process (SI Section \ref{Ssec:kinetics}). 
The predicted scaling laws with $t$ are confirmed in Fig.\ref{fig:n4}.



\begin{figure*}
\centering
\includegraphics{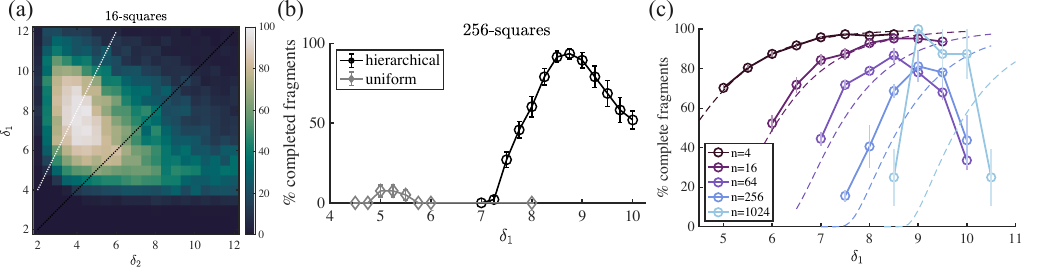}
\caption{Hierarchical assembly leads to high yield. (a) Building 16-squares with varying side energies $\delta_1,\delta_2$. Colors show the percentage of fully-completed 16-squares after $128\times 10^3$ MC sweeps. The average yield is nearly 100\% for a range of energies near $\delta_1 = 2\delta_2$ (white dotted line), while the yield is notably worse for identical interactions $\delta_1 = \delta_2$ (black dotted line). 
Simulations considered $N_c{=}10$ copies of a 16-square, and for each value of $(\delta_1,\delta_2)$ were repeated 12 times to obtain an average yield.
(b) Yield of perfect 256-squares in simulations of $N_c{=}4$ copies of a 256-square, for both hierarchical interactions ($\delta_1{=}2\delta_2{=}4\delta_3{=}8\delta_4$) and uniform interactions ($\delta_1{=}\delta_2{=}\delta_3{=}\delta_4$). The yield is plotted after $248\times 10^5$ MC sweeps, and simulations for each $\delta_1$ were repeated 12 times for hierarchical interactions, and 10 times for uniform interactions. 
(c) Yields of $n{=}4^k$-squares using hierarchical interactions (markers/solid lines), as a function of strongest interaction energy $\delta_1$, calculated at times increasing by a factor of 8 for each generation (except for $k{=}5$ as described in the text). 
Dashed lines correspond to equilibrium approximation Eq.\ref{yieldvsdelta2}.
All simulations consider 2048 monomers, so the number of copies is $N_c = 512, 128, 32, 8, 2$ for 4-,  16-,  64-,  256-, 1024-squares respectively. Systems of 64-, 256-, and 1024-squares are repeated four times and the yields are averaged. Vertical lines are one-standard deviation error bars.
}\label{fig:n16}
\end{figure*}

\subsection{Assembling 16-squares is best done using hierarchical interactions}

Since we can make 4-squares for a range of interaction energies, we now ask whether we can glue these 4-squares together in the same arrangement to form 16-squares, using the same overall interaction energies that we used for the monomers in the 4-squares. If we used interaction energy $\delta_1$ to make 4-squares, and we wish the overall interaction energy between 4-squares to also be $\delta_1$, then interactions between sides of monomers that are native to the 16-square but that are not in a 4-square must be $\delta_2 = \delta_1/2$ 
(Figure \ref{fig:hier}). 

We explore whether this intuition is correct by varying the interaction energies $\delta_1,\delta_2$ for a system of 16-squares (Figure \ref{fig:n16} left). 
The highest average yield after a given fixed time is 97.5\%, for $(\delta_1,\delta_2) = (8,4.5)$. The yield is not sensitive to these parameters: it is above  90\% for a selection of parameters roughly in the range $(\delta_1,\delta_2) \in [6.5,8.5]\times[3.5,5]$. This is near the line $\delta_2=\delta_1/2$,  validating our hypothesis that hierarchical interactions should lead to efficient assembly. 
It is notable that uniform interaction energies ($\delta_1{\approx}\delta_2$) never lead to such high yield over the time scales considered --  the highest uniform yield is 82.5\% for $\delta_1{=}\delta_2{=}6$, but most yields are much smaller.



\begin{figure*}
\centering
\includegraphics{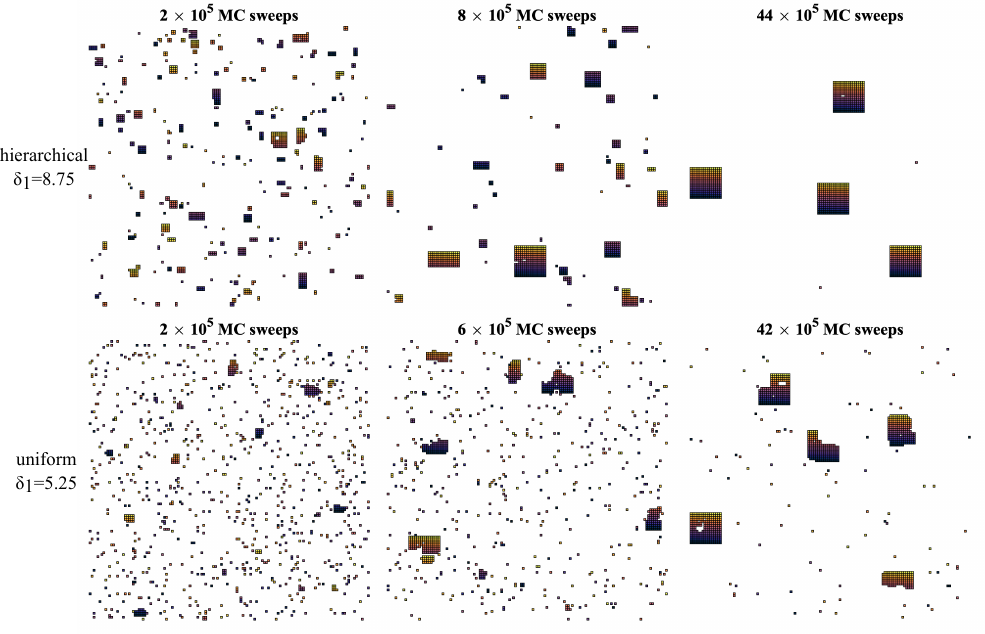}
\caption{Snapshots of individual simulations from Fig. \ref{fig:n16} (b) of assembling 256-squares at the optimal interaction energies for each: hierarchical ($\delta_1=8.75$, top) and uniform ($\delta_1=5.25$, bottom). Each panel shows the full simulation domain.
}\label{fig:n256traj}
\end{figure*}

\begin{figure}
\centering
\includegraphics{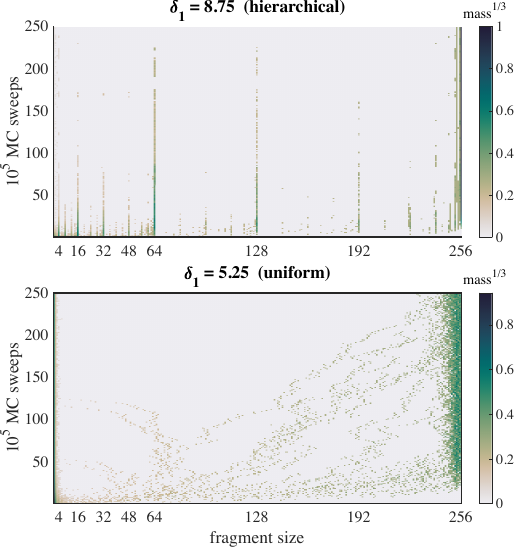}
\caption{The mass distribution of fragments of varying size ($x$ axis) as a function of time ($y$ axis) as indicated in color, for the simulations described in Fig. \ref{fig:n16} (b).  Parameters correspond to the best-assembling hierarchical energy ($\delta_1=8.75$, top), and the best-assembling uniform energy ($\delta_1=5.25$, bottom). 
For hierarchical interactions, most mass lies in fragments of sizes 4, 16, 64, and finally 256, with some mass also in multiples of these sizes. 
For uniform interactions, individual growth trajectories appear in the mass plot, consistent with the notion that clusters  grow by adding monomers one by one. The color scheme is mass$^{1/3}$ to highlight low-mass details.
}
\label{fig:n256mass}
\end{figure}

\subsection{Assembling $n$-squares with high yields via hierarchical interactions}

We now apply this principle of choosing interactions in a self-similar manner to build larger objects. The largest system we can explore systematically is a 256-square, which requires 
 additional side energies $\delta_3{=}\delta_2 / 2$ to make the intermediate 64-squares, and $\delta_4{=}\delta_3/2$ to make the final 256-squares, as illustrated in Fig.\ref{fig:hier}. The yield of 256-squares with hierarchical interactions is compared to the yield with uniform interactions in Fig.\ref{fig:n16} (right). The yield with hierarchical interactions reaches 95\% at late times with $\delta_1{=}8.75$, but is notably high over a broad range of timescales and a range  $\delta_1$ values spanning about 2$k_BT$; the yield is not sensitive to the particular choices of parameters. 
 In contrast, the maximum yield for uniform interactions over this timescale is only 8\%; this occurs near $\delta_1{=}5$ for a narrow range of interaction energies spanning about $0.5k_BT$. 
  
As a check that we can extend hierarchical assembly by one more generation we simulated  $N_c=2$ copies of a 1024-square using $\delta_1{=}10$, for a time of $1.15\times10^6$ MC sweeps, and repeated 4 times.  
 Out of the 8 total copies of a 1024-square simulated, 7 copies formed perfectly, and 1 copy was missing two monomers from its interior.  Snapshots of a perfectly-assembling system are shown in the SI,  Fig. \ref{fig:n1024}. 

Hierarchical assembly as the number of generations  $k$ varies is compared in Fig.\ref{fig:n16}(c). To choose the duration of these runs systematically, we consider $k=1$ and determine a time $t_1= 64\times10^3$  MC sweeps at which $y_1$ is close enough to equilibrium over a range of $\delta_1$ values. For larger $k$, we chose $t_{k} = 8^{k-1}t_1$, based on scaling time with the diffusion timescale for each generation, $\tau_D^k = (D_k\rho)^{-1}$ (for each subsequent generation, $\rho$ decreases by a factor of 4  for fixed volume fraction $\phi$, and $D_k$ decreases by a factor of 2).  Generation $k=5$ is  run for about half that time  because of computational limitations. 

Fig. \ref{fig:n16}(c)  shows that the yields  are high over a range of $\delta_1$-values for all generations, though they decrease slightly with each new generation as explained below. An exception is $k=5$ which has abnormally high yields, which we attribute to finite-size effects (Fig. S3; 
SI Section \ref{sec:finitesize}). 
The curves appear shifted to the right by about $1k_BT$ for each additional generation. This is consistent with \eqref{e1}, which shows that for fixed error rate $\Delta_1$ in the first generation and fixed volume fraction $\phi=4^k\rho$, the interaction energy must increase by about $3/4\log 4\approx 1.04k_BT$ for each additional generation (see also SI, Eq.\eqref{S:deltavsyield}). 

The yield curves must eventually go down with interaction strength, as already apparent for $k\geq 3$ for the range of interaction strengths considered.  
Upon visual inspection, we see small vacancy defects in target structures formed with larger $k$ and $\delta_1$. 
We will study systematically this point below.

\subsection{Dynamics of hierarchical assembly}

Fig.\ref{fig:n256traj} illustrates via snapshots the comparison between hierarchical (top) and regular (bottom) assembly. The difference is visually striking: for the hierarchical assembly, fragments often consist of completed meta-squares with flat boundaries. By contrast, regular self-assembly is more similar to crystal growth, where shapes with rough boundaries appear.

This difference in the assembly pathways can be further probed by studying the mass distribution in fragments of different sizes as a function of time (Figure \ref{fig:n256mass}). For hierarchical interactions, most mass at early times is concentrated in fragments of size 1, then 4, then 16, then 64, with most remaining mass  in multiples of these sizes (8,12, 32, 48, 128, 192). 
For uniform interactions, the system mass is concentrated in fragments of size 1, and of sizes near but slightly less than 256, with very little mass in between. Individual growth trajectories in the mass plots resemble random walks in the space of fragment size, with a bias toward larger sizes. This feature is not present with hierarchical interactions. 

To quantify this behaviour further, in the  left panel of Fig.S2 
the mass in fragments of size $n=4,16,64,256$ is shown as a function of time, for a system of assembling 256-squares with optimal hierarchical interactions.
The peaks in the masses occur at subsequently increasing times and then decay, except for the targeted structure $n=256$. Noticeably, the decaying parts of the curves appear simply shifted in time by $1/\sqrt{n}$ as $n$ increases (right panel), consistent with scaling the decay with the diffusion timescales as we rationalize below.

\section{Constraints on hierarchical assembly}

We have shown that hierarchical assembly pathways may be created by choosing appropriate interaction strengths, and they can lead to efficient, high-yield assembly. We now consider the conditions under which such assembly will be successful, considering both thermodynamic and kinetic constraints.

\subsection{Themodynamics constraints} \label{sec:thermotheory}
We seek to estimate the yield $y_k$ of hierarchical self-assembly with $k$ generations,  as a function of the monomer density $\rho$ and of the interaction strengths. Let us denote by $\tilde \delta_j=2^j \delta_j$ the effective interaction between two fully formed squares at the $j^{th}$ intermediate generation, each consisting of $4^j$ monomers. We will assume that $\tilde \delta_j= \delta$ independently of $j$. We denote by $y_{j}\e{k}$ the yield of such squares. At the first generation, assuming that weaker interactions do not disturb the kinetic of the assembly process (see below), from Eq.\ref{e1} we get $y_{1}\e{k}=1-\Delta_1(\rho)$. We can now proceed recursively,  assuming 
that (i) the formation of  squares at generation $j$ is not affected by that of larger structures and (ii) that it only depends on the concentration of $j-1$ squares obtained in the process, which only depends on the yields of previous stages $j'$ with $1\leq j' \leq j$.
The yields of intermediate generations are thus approximated as
\begin{equation}\label{yieldvsdelta2}
y_{j}\e{k}(\rho) = y_{j-1}\e{k}(\rho)y_1\e{k}(\rho y_{j-1}(\rho)),
\end{equation}
and the target yield $y_k = y_{k}\e{k}$ is computed recursively. 
This yield approximation is tested against simulation data in Fig.\ref{fig:n16}(c) (dashed lines). It correctly captures the trend for the yield, and even agrees quantitatively for weak interactions and small $k$. 

Since a final yield of order unity requires a high yield already at the first generation, Eq.\ref{e1} implies
\begin{equation}
\label{e5}
e^{\delta}\gg\rho^{-3/4}.
\end{equation}
Thus the thermodynamic yield increases with  $\delta$. However, for large enough $\delta$  kinetic effects drastically reduce the yield, as we now discuss.

\subsection{Kinetic constraints}

{\it Bonds must have time to detach to avoid misfits:} The  assembly time must be larger than the time it takes any interaction to detach. Since the unbinding timescale for generation $j$, $\tau_B^j\sim e^{\delta}/D_j$, is largest for the final generation, this implies:
\begin{equation}
\label{e6}
t\gg e^{\delta}/D_k.
\end{equation}

{\it Remaining hierarchical:} For hierarchical assembly to proceed, interactions needed to build the generation $j+1$ should not interfere with the assembly of generation $j$; they should start to play a role only when the $j$-squares are fully formed. Incomplete substructures (such as triangles) of two adjacent $j$-squares  bind with an interaction $\delta/2$. They will thus stick together even when the $j$ squares are not completed, and remain attached for some time $\sim e^{\delta/2}/D_j$. If this time is much larger than the diffusion time $\tau_D^j\sim 1/(\rho D_j)$ for $j$-squares to meet, then partially formed structures spend most of their time bound to other partially formed structures. Proper hierarchical assembly can only happen when this does not occur,  implying:
\begin{equation}
\label{e4}
e^{\delta/2}\ll \frac{1}{\rho}.
\end{equation}

{\it Avoiding defects:} As discussed for the formation of 4-squares, the concentration of triangles will be abundant at any level $j$ at some point during the hierarchical assembly if the ratio of unbinding time over diffusion time $f^j\equiv \tau^j_B/\tau^j_D= e^{\delta} \rho$ is large. If the associated condition $f^j\ll 1$ or equivalently:
\begin{equation}
\label{e7}
e^{\delta}\ll \frac{1}{\rho}
\end{equation}
does not hold, we expect to see  defects composed of a triangle binding to three completed squares, such that an inaccessible hole is present in the interior of this structure. Indeed, the energy of removing the triangle to heal such a defect is precisely $\delta$; condition Eq.\ref{e7} ensures that the healing time is smaller than the characteristic time where intermediate structures meet and bind.  

Interestingly, if we denote by $t_j$  the time scale to form generation $j$, we have that $t_j$  is always inversely proportional to $ D_j$.  Indeed, $t_j$ corresponds to the slowest time between the unbinding time $\tau^j_B$ and the diffusion time $\tau^j$, and both satisfy this property. Since $D_j\sim 1/\sqrt{4^j}$, one expects that during the hierarchical assembly, subsequent generations should take 2 times longer to form. Thus, $t_j\sim 2^j\sim \sqrt{n}$, as confirmed in Fig.S2. 

\begin{figure*}
\centering
\includegraphics{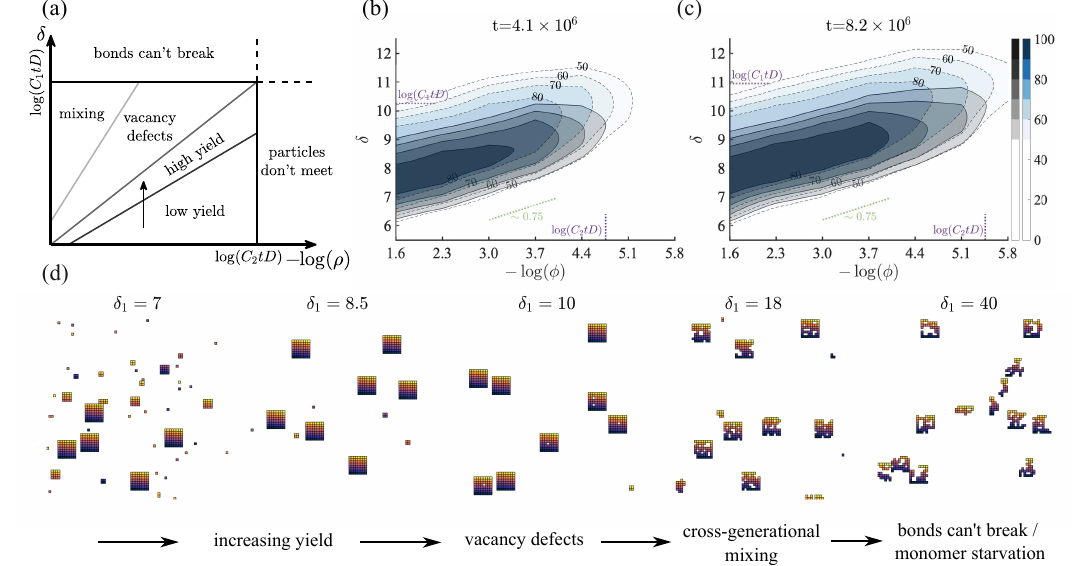}
    \caption{Toward a phase diagram for hierarchical self-assembly. (a): A schematic phase diagram, showing regions with different qualitative behaviour. The regions are demarcated by lines of the form (for increasing  $\delta)$ \eqref{e5}, \eqref{e7}, \eqref{e4},\eqref{e6}. (b,c): Contour plots of the yield for $n=64$ with hierarchical interactions as a function of strongest interaction energy $\delta$ and volume fraction $\phi$. The yield includes structures with 0 defects (grey shades / solid lines) and 2 defects (blue shades / dashed lines), at times (b) $t=4.096\times 10^6$ and (c) $t=8.192\times 10^6$ MC sweeps. Purple dotted lines show $\log (CtD)$ with $C_1=1/9$, $C_2=(1/35)/64$, and $D=1/16$ the diffusivity for a 16-square. Green dotted line has slope $0.75$ and is a guide to the eye.
    (d): Snapshots showing qualitatively how hierarchical assembly behaves as $\delta$ increases, for $n=64$ at fixed volume fraction $\phi=0.05$, at time $t=8.192\times 10^6$ MC sweeps.  
    Each panel is a portion of one simulation described in Fig. S4
    with $\phi=0.05$. 
    }
    \label{fig:phase}
\end{figure*}

\subsection{Phase diagram}
Our results are summarized in the qualitative phase diagram of Fig. \ref{fig:phase} (a). For fixed $\rho$, at low $\delta$ the yield is poor  as expected thermodynamically, and improves  as $\delta$ increases. The yield becomes large when $e^{\delta} > C/\rho^{3/4}$ as in Eq.\eqref{e5} (here $C$ represents an unknown constant). For $e^{\delta}> C/\rho$ as in Eq.\eqref{e7}, defects can start to appear. For even stronger interactions  $e^{\delta}> C/\rho^2$ as in Eq.\eqref{e4}, the notion of hierarchical assembly is lost, as interactions associated with different levels can bind concomitantly. Finally for $e^{\delta}>C\,tD$ as in Eq.\eqref{e6}, low-level bonds have no time to detach: kinetic traps appear already at small scales. Snapshots from different regions of the diagram showing these different qualitative behaviours are shown in Fig. \ref{fig:phase}(d).

Overall, from these arguments we expect that  yield is governed by a trade-off between thermodynamics, which favors large $\delta$, and defect formation, so that optimality is reached near the regime where defects appear, i.e. for $e^{\delta}$ not too large in comparison with $1/\rho$. 
However, at that stage structures are still formed, but with a few defects. Thus, if a more relaxed definition of yield that includes structures with a small number of defects is considered, we expect  that this optimal relaxed yield is obtained for larger $\delta$.

To test these  predictions, we simulate systems of $n=64$-squares on intermediate time scales to obtain an empirical phase diagram, as described in more detail in the SI (Section \ref{sec:n64}). 
Fig. \ref{fig:phase} (b,c) show contour plots of the yield (grey shades) as a function of $(-\log\phi,\delta)$ (recall $\phi = 64\rho$), for two assembly times separated by a factor of 2.  We superimposed contour lines for a relaxed definition of yield (blue shades), which includes structures with at most 2 defects (hence which are at least 97\% complete).  Key observations include:
\begin{itemize}
    \item As predicted, the high-yield region disappears when $\delta$ is too small or too large, and when the volume fraction $\phi$ is too small.  
    
    \item Both the threshold inverse packing fraction $1/\phi_c$ and the threshold $\delta_c$ beyond which the yield degrades  are consistent with our predictions  $\log(CtD)$, where $C$ is a constant that depends on the quantity considered. This is  apparent when comparing panels (b) and (c) of Fig. \ref{fig:phase}. Specifically,  we determine from the middle plot  (b) the constants $C_1,C_2$ such that $\delta_c=\log(C_1 tD)$ and $\log (1/\phi_c)=\log(C_2 tD)$, where $\delta_c$ and $1/\phi_c$ are defined as the largest values of $\delta$ and $1/\phi$ for which 50\% yield is reached. We can see that  our predictions for $\delta_c$ and $1/\phi_c$ still hold  as the time $t$ is changed in panel (c).

    \item As predicted, both definitions of yield initially increase at the same rate with $\delta$. However, the relaxed definition of yield that allows for defect remains high for a wider range of $\delta$ values, validating our prediction that the assembly starts to degrade due to vacancy defects. 

     \item The slopes of the lines where high yield appears are $\approx 0.6{-}0.7$, close to the prediction of Eq.\ref{e5} of slope $0.75$. We measure similar slopes for the emergence of defects (corresponding to suppression of the high yield region, while the relaxed definition is still high), close to the prediction of slope 1, Eq.\ref{e7}.
 
\end{itemize}
The emergence of mixing  \eqref{e4} corresponds to the decrease of the relaxed definition of yield. Measuring it is difficult, as it requires to be in the  regime of large $n$ and $t$ which hard to access in simulations; see the discussion in the SI Section \ref{sec:n64}. 

\subsection{Role of non-native interactions}
The yield is expected to degrade when there are non-native interactions present, also called crosstalk.  
 Crosstalk has been shown to significantly degrade the yield of self-assembly once the average crosstalk energy is above a threshold \cite{murugan2015undesired}, and some amount of crosstalk is inevitable in physical systems \cite{Huntley.2016}. For hierarchical assembly, we expect that cross-talk will affect mostly the assembly of the largest structures (i.e. the squares of size $n/4$). If the cross-talk energy is a constant $U$, these structures will bind non-specifically with an energy $\sim U \sqrt{n}$ (neglecting numerical pre-factors), thus the threshold cross-talk energy $U_c(n)$ beyond which self-assembly is impossible follows $U_c \sim 1/\sqrt{n}$. If the cross-talk energies are of mean zero and standard deviation $\sigma$, then problems are milder and can be estimated with similar arguments. In particular,  the threshold for assembly expected to scale as $\sigma_c\sim n^{-1/4}$.

 To check the effect of crosstalk we performed simulations for different $n$-squares with different levels of crosstalk, assumed to be of constant magnitude (details in SI Section \ref{sec:crosstalk}). 
 For each $n$, the yield was unaffected for crosstalk energy below a threshold $U_c(n)$, beyond which it decreased rapidly  (Fig.S6). 
 The threshhold 
 follows approximately our prediction $n^{-1/2}$. 

\section{Conclusion}

The dominant protocol for high yield addressable self-assembly is based on heterogeneous nucleation: start from high temperature where nucleation is rare so that forming structures do not compete, and eventually lower temperature to make the structure form with fewer defects.  Biology however offers examples of self-assembly with high yield at a fixed temperature, raising the possibility  that other interaction designs may display  higher yields. We have introduced a simple model to illustrate that hierarchical self-assembly is a  powerful design principle for the interaction pattern, leading to very high yield. In this case, the dynamics is markedly different from heterogeneous nucleation. Particles first assemble into sub-units, themselves forming units and so on in a sequential manner, instead of displaying structures whose masses continuously grow. 

We have established that for hierarchical self-assembly, various quantities  scale with the mass $m$ of sub-structures considered.  On the one hand, 
pairwise interaction should decrease at the boundary of large units, such that structures at a given scale display an overall interaction independent of that scale. For compact sub-structures  that fill up space, this implies a microscopic interaction strength decreasing  as $m^{(d-1)/d}$ where $d$ is the spatial dimension. 
We argued that the optimal interaction strength $\delta$ will lie in some interval, with a lower bound given by thermodynamic considerations, and an upper bound determined when the diffusion time becomes slower than the unbinding time and hence vacancy defects start to appear. 
Determining quantitatively the precise location where yield is optimal for a given assembly duration remains a theoretical challenge for the future.


Overall, we have shown that hierarchical assembly is a highly successful strategy in a simple model of sticky squares, but we expect it to be a useful principle in  more general situations. Future work can explore how to extend these ideas to systems with different geometries, including to DNA bricks where our predictions could be 
tested experimentally. 


Our analysis also suggests to revisit the role of hierarchical assembly in protein folding. Folding is believed to  occurs successfully because the energy landscape presents a funnel shape \cite{bryngelson1995funnels}, a description that however does not  distinguish between very distinct folding mechanisms. Multiple mechanistic views exist, ranging from pure nucleation \cite{dokholyan2000identifying} to hierarchical folding \cite{baldwin1999protein,dill2008protein} as well as intermediary scenarios including both phenomena acting at different scales \cite{fersht1997nucleation}. Studying this question would require extending our analysis to the case where monomers come from a finite set, and are constrained to form a chain. 
Interestingly, this problem may become experimentally accessible in soft matter systems,   for which chains of particles with specific interactions can be built \cite{mcmullen2022self}.

\begin{acknowledgments}
We thank J. Brujic, M. Johnson, D. Gracias, A. Grosberg and R. Schulman for discussions.  M.W acknowledges support from the Simons Foundation Grant (No. 454953 Matthieu Wyart). 
M.H.C. acknowledges support from the Alfred P. Sloan Foundation, and from the Natural Sciences and Engineering Research Council of Canada (NSERC), RGPIN-2023-04449 / 
Cette recherche a été financée par le Conseil de recherches en sciences naturelles et en génie du Canada (CRSNG). 
\end{acknowledgments}

%

\bibliography{HierarchicalAssembly}

\bigskip
\hrulefill\\

\onecolumngrid


\appendix
\makeatletter
\def\@seccntformat#1{\@ifundefined{#1@cntformat}%
   {\csname the#1\endcsname\space}
   {\csname #1@cntformat\endcsname}}
\newcommand\section@cntformat{\appendixname\thesection.\space} 
\makeatother
\renewcommand{\thesection}{S\arabic{section}}
\counterwithin{equation}{section}
\counterwithin{figure}{section}
\counterwithin{table}{section}

\section*{SUPPLEMENTARY INFORMATION}


\section{Equilibrium calculations, details}\label{sec:theory}

This section contains a few details supporting the calculations in Section \ref{sec:thermotheory} of the main text. 

First suppose we wish to make a 4-square, and consider all the intermediate fragment types:   there are 4 types of monomers with energy 0, 4 types of 2-mers with energy $-\delta$, 4 types of trimers with energy $-2\delta$, and 1 type of 4-mer with energy $-4\delta$. Let $n_i$ be the total number of $i$-mers observed, and let $c_i$ be the concentration of each type of $i$-mer, so that $c_1 = n_1/4V$, $c_2 = n_2/4V$, $c_3 = n_3/4V$, $c_4 = n_4/V$. 
The different species interact with reactions
\begin{align}
\text{1-mer}+ \text{1-mer} &\leftrightarrow \text{2-mer}, \nonumber\\
\text{1-mer}+\text{2-mer} &\leftrightarrow \text{3-mer}\nonumber\\
\text{2-mer}+\text{2-mer}&\leftrightarrow \text{4-mer}, \nonumber\\
\text{1-mer} + \text{3-mer} &\leftrightarrow \text{4-mer}.
\label{c4rxns}
\end{align}
The equilibrium concentrations, assuming equal probabilities of each type of $k$-mer, and small $\phi$ so the entropy of $\alpha$ units of $i$-mers is $\approx \alpha\log c_i$, must therefore satisfy
\begin{align}
2\log c_1 &= \log c_2 - \delta, \nonumber\\
\log c_1 + \log c_2 -\delta &= \log c_3 - 2\delta  \nonumber\\
2\log c_2 - 2\delta &= \log c_4 - 4\delta, \nonumber\\
\log c_3 + \log c_1 - 2\delta &= \log c_4 - 4\delta  \nonumber 
\end{align}
Solving gives 
\begin{equation}
    c_1 = e^{-\delta}c_4^{1/4}, \qquad c_2 = e^{-\delta}c_4^{1/2}, \qquad c_3 = e^{-\delta}c_4^{3/4}. 
\end{equation}
The equilibrium equations are completed using conservation of mass, $(4c_1 + 4\cdot 2c_2 + 4\cdot 3c_3 + 4c_4)V = 4N_c$, or
\begin{equation}\label{S:c4mass}
4c_1 + 8c_2 + 12c_3 + 4c_4 = \phi,
\end{equation}
which after substituting for $c_1,c_2,c_3$ leads to 
\begin{equation}\label{S:c4}
c_4^{1/4} + 2c_4^{1/2} + 3c_4^{3/4} + c_4e^{\delta} = e^{\delta}\phi/4.
\end{equation}
This equation may be solved numerically for $c_4$, given $\delta,\phi$. We develop an approximation valid for high yield (large $c_4$) momentarily. 

Now suppose we wish to form a $4^k$-square using hierarchical interactions. The system must first form 4-squares; therefore we must similarly ask which value of $\delta=\delta_1$ makes these 4-squares have high yield, \emph{before} we turn on the hierarchical interactions (i.e. assuming $\delta_2=\delta_3=\cdots=0$). The difference from a system of pure 4-squares is that now there are $4^{k-1}$ distinct copies of each 4-square, so the equation for mass conservation \eqref{S:c4mass} is replaced by 
\begin{equation}\label{S:c4kmass}
4c_1 + 8c_2 + 12c_3 + 4c_4 = \frac{\phi}{4^{k-1}}
\end{equation}
because the overall concentration of each type of monomer is reduced by a factor of 4. 
Therefore, the concentration $c_4=n_4/4^{k-1}V$ satisfies \eqref{S:c4}, but with $\phi\to\phi/4^{k-1}$. 

When the yield $y_1\e{k} = 4^kc_4/\phi$ is high, we may obtain a perturbative solution to \eqref{S:c4kmass} by letting $y_1\e{k} =  1-\Delta_4\e{k}$ with $\Delta_4\e{k}\ll 1$. 
%
At low density  where $c_4 \ll 1$, we must have 
$c_1\gg c_2\gg c_3$, so we drop $c_2,c_3$. 
Because of the high yield assumption, $c_4 \gg c_1$, so we only include the perturbation in $c_4$, making the approximations $c_4 \approx (1-\Delta_4\e{k})\phi/4^k$, $c_1\approx (\phi/4^k)^{1/4}$. 
Solving \eqref{S:c4kmass} using these approximations gives 
\begin{equation}\label{S:yieldvsdelta}
\Delta_4\e{k} \approx e^{-\delta}\left(\frac{\phi}{4^k}\right)^{-3/4}.
\end{equation}
This allows us to approximately solve for the interaction energy required to obtain a desired yield: 
\begin{equation}\label{S:deltavsyield}
\delta \approx k\frac{3}{4}\log 4 - \frac{3}{4}\log \phi - \log \Delta_4\e{k}. 
\end{equation}


\section{Kinetics of forming a 4-square}\label{Ssec:kinetics}

We may solve for the time-dependent concentrations of 4-squares by solving nonlinear ODEs corresponding to the different sub-reactions involved in forming a 4-square. 
We consider the reactions  \eqref{c4rxns}, and assume (i) infinite volume at fixed volume fraction (ii) the system is well-mixed, and (iii)  the initial concentrations of each sub-type of fragment are the same (monomer, dimer, trimer). Then standard methods give the system of reaction equations
\begin{align}
\dd{c_1}{t} &= -2k_{11}c_1^2 + 2k_{2}^{\rm off}c_2 - 2k_{21}c_1c_2 + 2k_{3}^{\rm off}c_3 - k_{31}c_3c_1 + k_{43}^{\rm off}c_4\nonumber\\
\dd{c_2}{t} &= k_{11}c_1^2 - k_2^{\rm off} c_2 - 2k_{21}c_1c_2 + 2k_3^{\rm off}c_3 - k_{22}c_2^2 + k_{42}^{\rm off} c_4\nonumber\\
\dd{c_3}{t} &= 2k_{21}c_1c_2 - 2k_{3}^{\rm off} c_3 - k_{31}c_3c_1 + k_{43}^{\rm off} c_4  \nonumber\\
\dd{c_4}{t} &= 4k_{31}c_3c_1 - 4k_{43}^{\rm off} c_4 + 2k_{22}c_2^2 - 2k_{42}^{\rm off} c_4. 
\label{concode}
\end{align}
Here $k_{(\cdot)}$ represent binding rates and $k^{\rm off}_{(\cdot)}$ represent unbinding rates. 
These equations can be shown to conserve total mass, $4c_1+8c_2+12c_3+4c_4$. 

We choose the binding rates to be proportional to the sum of the species' diffusion coefficients: 
\begin{align*}
k_{11} &= k_0\times2D_1 & (1+1 \to 2)\\
k_{21} &= k_0\times(D_1+D_2) & (2+1\to 3)\\
k_{22} &= k_0\times2D_2  &(2+2 \to 4)\\
k_{31} &= k_0\times(D_1+D_3) & (3+1 \to 4)
\end{align*}
where $D_1$ is the diffusion coefficient of a monomer, and $D_2 = D_1/(3/2)$ is the diffusion coefficient of a 2-mer in the wide direction, and $D_3 = D_1 / (1+\sqrt{2}/3)$ is the diffusion coefficient of a 3-mer, in both directions. 
The base rate (if comparing with simulations) is $k_0 = 0.25$; this accounts for the directionality of the interactions, since particles can only bind if they collide in one of 4 possible orientations. 

The unbinding rates are chosen to satisfy detailed balance, as
\begin{align*}
k_2^{\rm off} &= k_{11}e^{-\delta} & (2\to 1+1)\\
k_3^{\rm off} &= k_{21}e^{-\delta} & (3 \to 1+2)\\
k_{42}^{\rm off} &= k_{22}e^{-2\delta}& (4 \to 2+2)\\
k_{43}^{\rm off} &= k_{31}e^{-2\delta} & (4 \to 3+1)
\end{align*}

We solved \eqref{concode} numerically to find $\delta_{\rm opt}(t)$, the interaction parameter $\delta_1$ that maximizes the concentration $c_4(t)$ at fixed $t$, and the corresponding value of the optimal yield, $y_4^{\rm opt}(t) = c_4(t)/\rho$, at fixed volume fraction $\phi=0.05$. 
We performed the computations in Julia using packages DifferentialEquations, SciMLSensitivity, OptimJL, to solve the equations numerically, compute the gradient of $c_4(t)$ using automatic differentiation, and find the optimal parameter values using the BFGS algorithm. 
Fig. \ref{fig:n4} shows a clear scaling of the optimal energy $\delta_{\rm opt}(t) \sim t^{0.91}$, and of the optimal yield $1-y_4^{\rm opt}(t) \propto t^{-0.97}$. These scalings are consistent with our predictions from \eqref{e2},\eqref{e3}.


\section{Additional figures and data}

This section collects additional figures and data to support our claims in the main text. 

\subsection{A system of $n=1024$-squares assembling}
\begin{figure}
\centering
\includegraphics{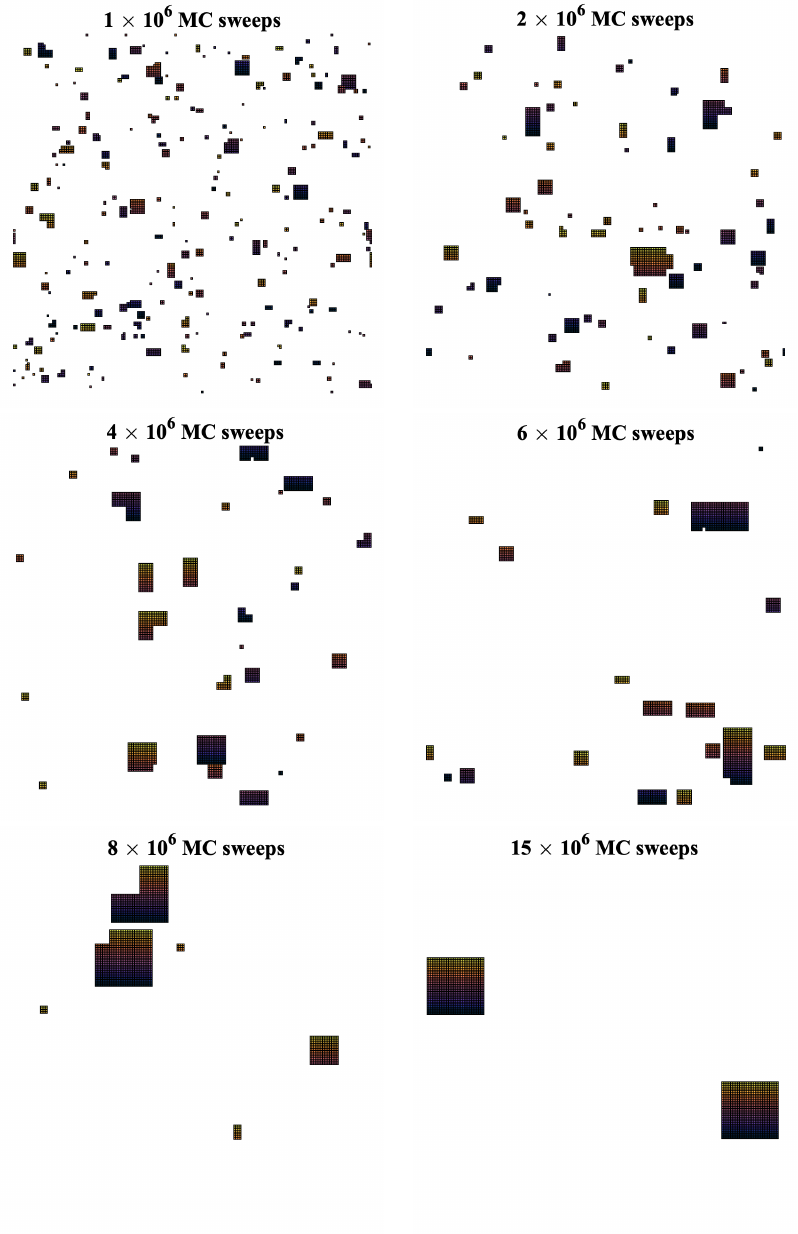}
\caption{Snapshots of $N_c=2$ copies of a 1024-square assembling, using hierarchical interactions with $\delta_1=10$. Each panel shows the full simulation.
}\label{fig:n1024}
\end{figure}

Figure \ref{fig:n1024} shows snapshots of a system of 1024-squares assembling with hierarchical interactions. Computational limitations mean we can only consider 2 copies of this system, however it is notable that we achieve very high assembly yields. 

\subsection{Mass versus time}

\begin{figure}
\centering
\includegraphics[width=0.45\linewidth]{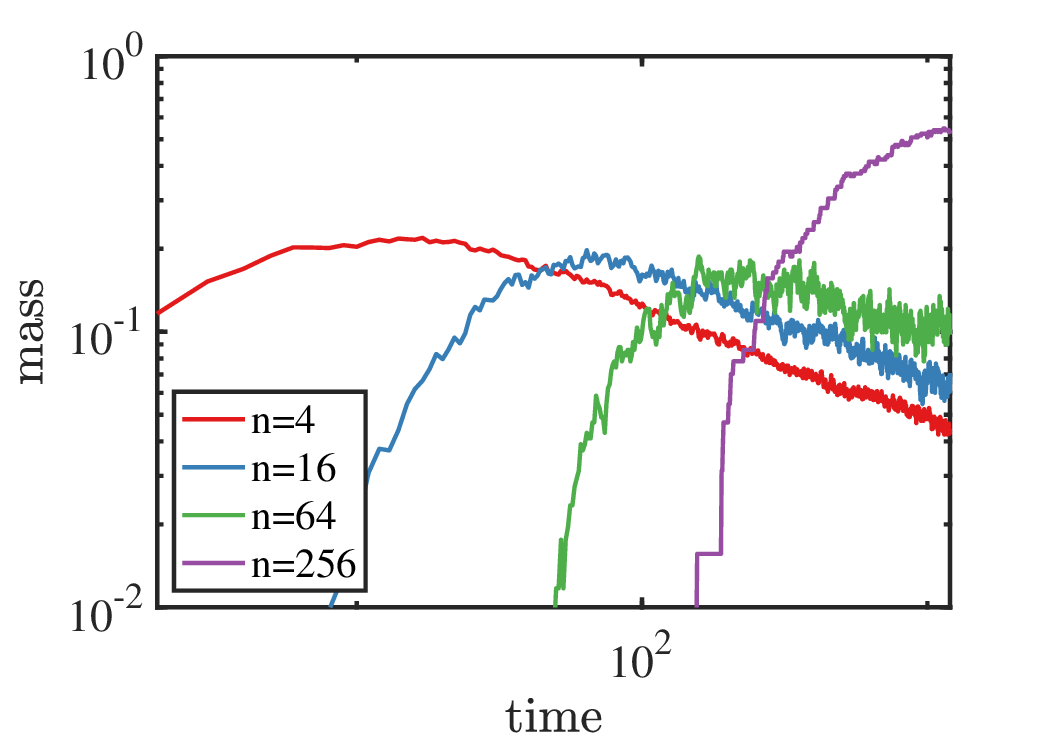}
\includegraphics[width=0.45\linewidth]{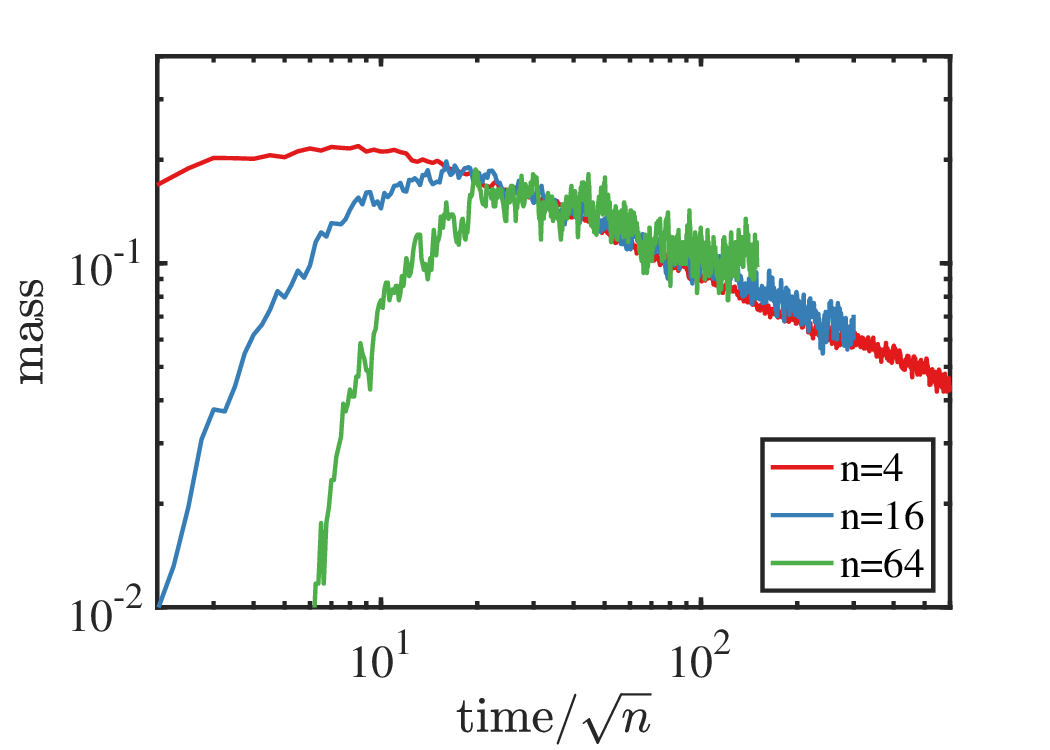}
\caption{(Left) Fraction of the system mass in fragments of size $n=4,16,64,256$, as a function of time, for a system of $N_c{=}16$ 256-squares with $\delta_1=8.75$, repeated 8 times. Time is measured in units of $10^4$ MC sweeps. (Right) Same data but with time rescaled by $1/\sqrt{n}$. }\label{fig:masstime}
\end{figure}

Figure \ref{fig:masstime} shows the fraction of the system's mass in fragments of size $n=4,16,64,256$, as a function of time, for a system of 256-squares. 
The first plot shows the raw data, while the second rescales time by $t\to t/\sqrt{n}$. The decay rates align for this rescaling, supporting our claim that  subsequent generations in this hierarchical assembly process should take 2 times longer to form.


\subsection{Finite size effects}\label{sec:finitesize}

\begin{figure*}[h]
\centering
\includegraphics[width=0.6\linewidth]{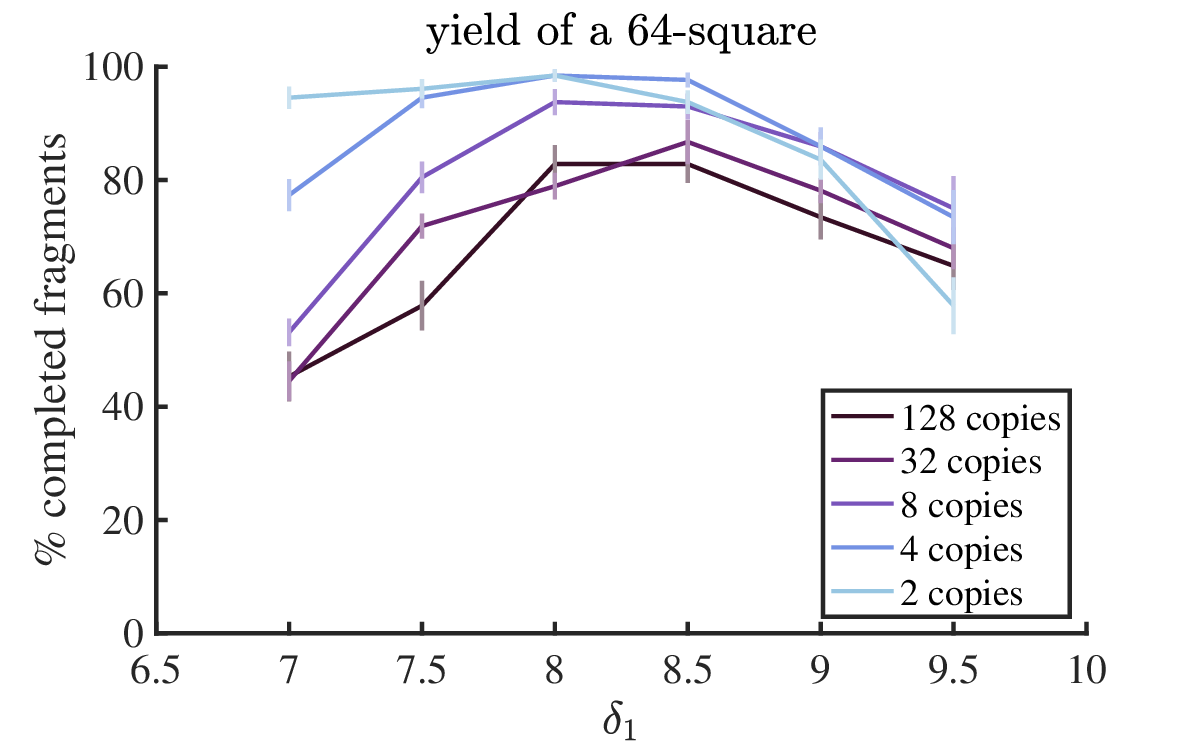}
\caption{Testing finite size effects in simulations of a 64-square with hierarchical interactions, at various values of $\delta_1$. The simulations consider $N_c$ copies of a 64-square, with values of $N_c$ shown in the legend, and with statistics averaged over $N_r$ independent realizations chosen such that there are a total of $N_cN_r = 128$ total copies simulated for each choice of parameters. Vertical lines are one-standard deviation error bars.  
}
\label{fig:finitesize}
\end{figure*}

We tested  the effect of simulating a finite number of copies of each target square. 
We ran simulations of $N_c$ copies of a 64-square with hierarchical interactions and with different values of $N_c$, repeated $N_r = 128/N_c$ times to obtain sufficient statistics. The yields are shown in Figure \ref{fig:finitesize}. This figure shows that finite-size effects tend to increase the yield, especially at the smallest values of copy size, $N_c \leq 4$, where the yield is significantly higher for small $\delta_1$. For $N_c=2$ with larger $\delta_1$, the yield drops steeply.


\subsection{Simulations for $n=64$ to construct the phase diagram}\label{sec:n64}


\begin{figure*}\centering
  \includegraphics[width=\linewidth]{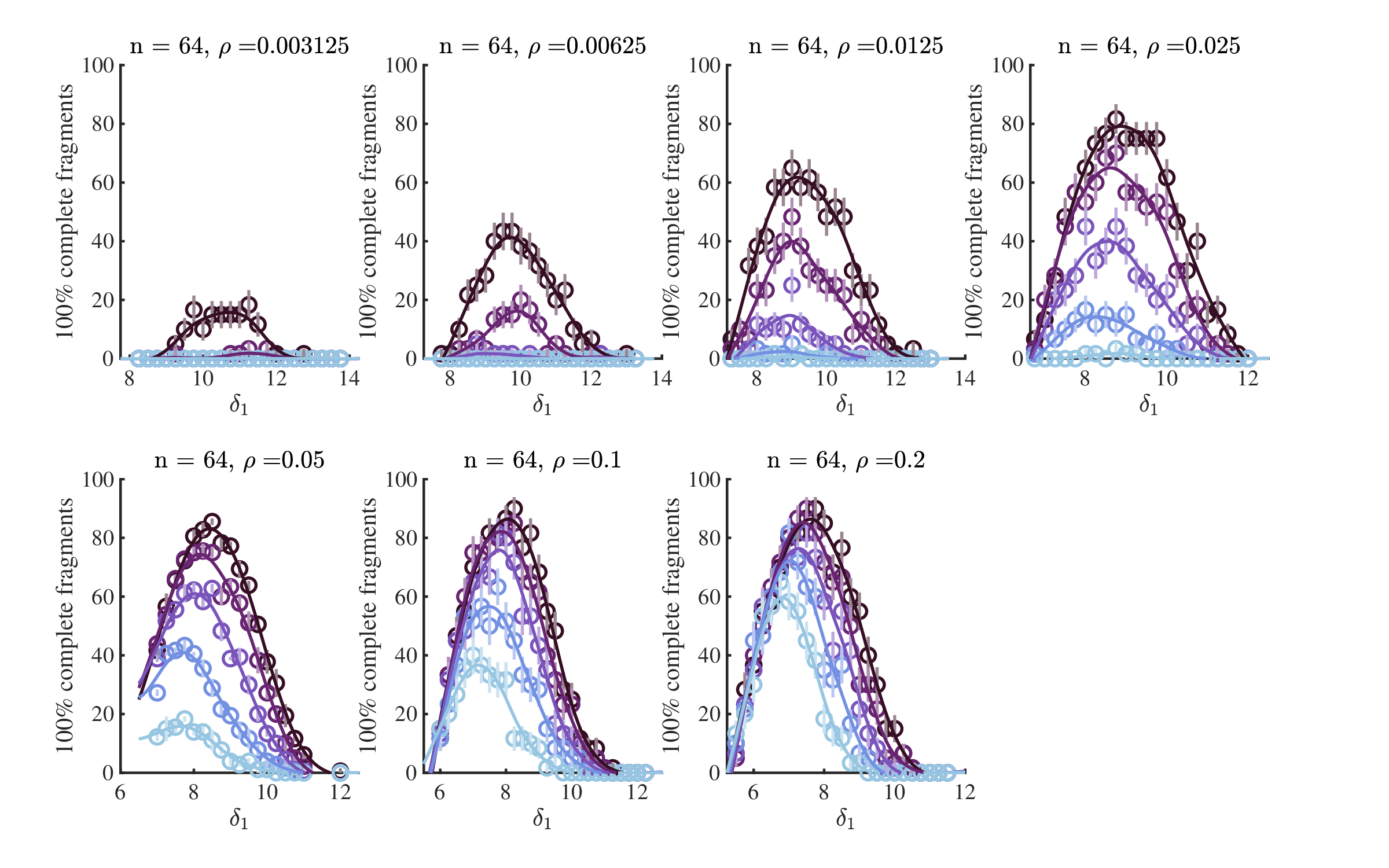}
\caption{Perfect yield from simulations of $n=64$-squares at different volume fractions $\phi$, with a range of values of $\delta_1$. Each marker corresponds to a simulation of $N_c=60$ copies of a 64-square. All simulations are performed once, except for $\phi=0.05 $ which is performed 3 times. Vertical lines are one-standard-deviation error bars, estimated as for a Bernoulli random variable (for all data except $\phi=0.05$), or from the (unbiased) standard deviation over the 3 independent simulations, for $\phi=0.05$. 
Curves of different colours show the yield at times increasing by a factor of 2, with the darkest colour corresponding to the maximum time $t=8.192\times 10^6$ MC sweeps. Solid lines correspond to a spline fit through the data. }\label{fig:yield100}
\end{figure*}

\begin{figure*}\centering
  \includegraphics[width=\linewidth]{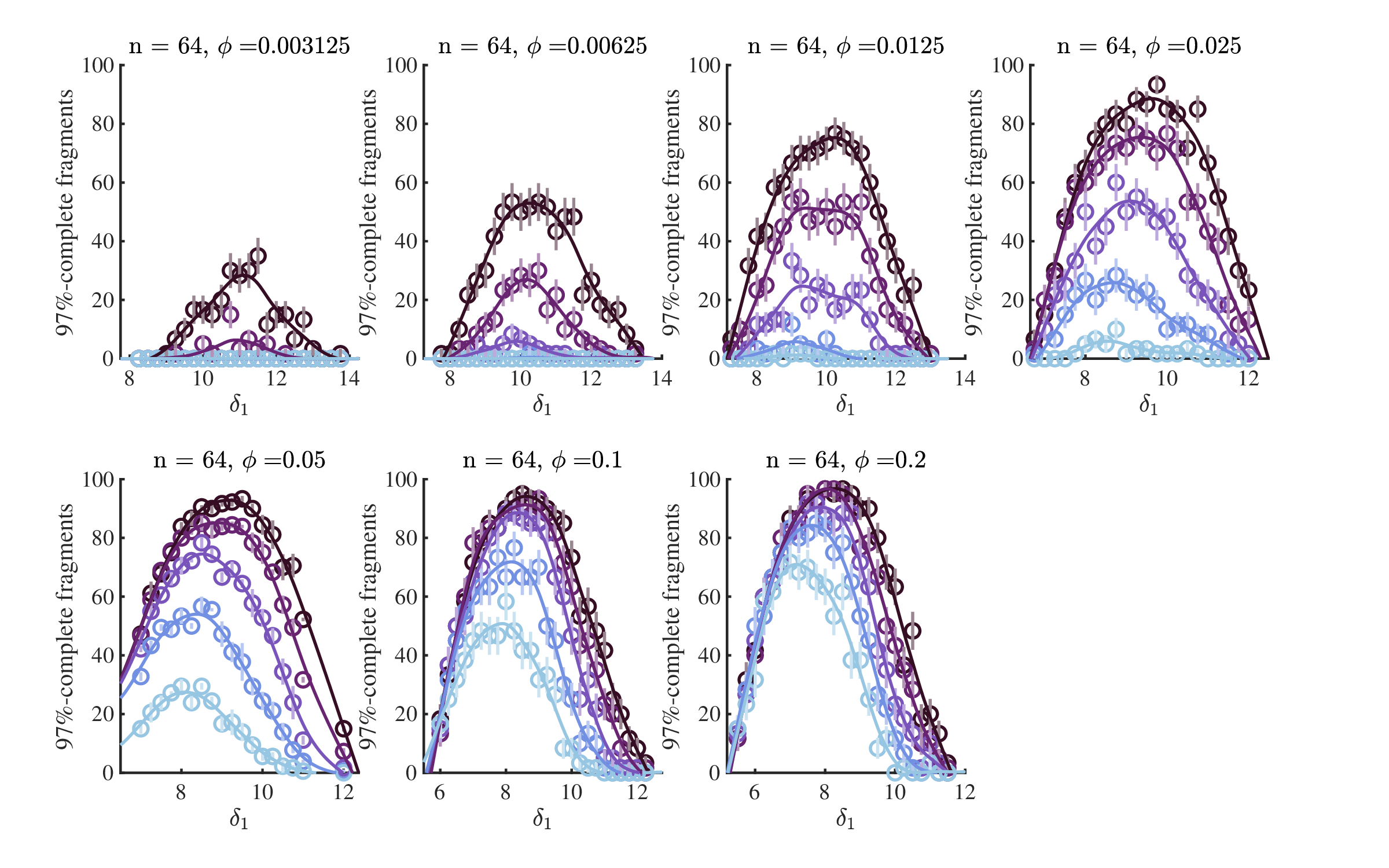}
\caption{Same as Figure \ref{fig:yield100}, but here we are showing the ``relaxed'' yield, corresponding to structures that are at least 97\% complete (contain at most 2 missing particles).}\label{fig:yield97}
\end{figure*}

We performed extensive simulations of $64$-squares assembling with hierarchical interactions at different densities, over a range of values of $\delta_1$. The yield data is shown in Figure \ref{fig:yield100} (yield with 0 defects) and Figure \ref{fig:yield97} (yield including structures with 2 defects). 
To smooth out the statistical fluctuations, we fit a spline (solid curves) through the data at each volume fraction, using MATLAB's \texttt{smoothingspline} function with smoothing parameter 0.95, and used the fitted curves to construct the phase diagram in Figure \ref{fig:phase}. 

Note that a quantitative departure with our predictions is  apparent in Fig.\ref{fig:phase}: the slopes of the contour lines delineating high yield are similar, and fitting a line yields values around 0.6-0.7. Concerning the lower contour line, this value is close to the slope 3/4 predicted by the criterion for thermodynamic yield of Eq.\eqref{e5}. Concerning the upper contour line, the observed slope is significantly smaller than the prediction for the onset of vacancy defects, Eq. \eqref{e7}, predicting a slope of 1, and the constraint of hierarchical self-assembly of Eq.\eqref{e4}, predicting a slope 2. 
Testing quantitatively this latter asymptotic prediction is however difficult, as it applies in the limit of  large assembly size $n$ and time $t$ that is hard to access in simulations. If $n$ is too small (the assembly is not hierarchical enough), annealing of poorly assembled structures will start to play a role and improve the yield. Moreover, testing Eq.\eqref{e7},\eqref{e4}  requires that the value of $\delta$ it predicts is much smaller than $\log(tD)$ where a cross-over occurs toward a regime where individual bond cannot break.


\subsection{Crosstalk simulations}\label{sec:crosstalk}

\begin{figure*}\centering
  \includegraphics[width=\linewidth]{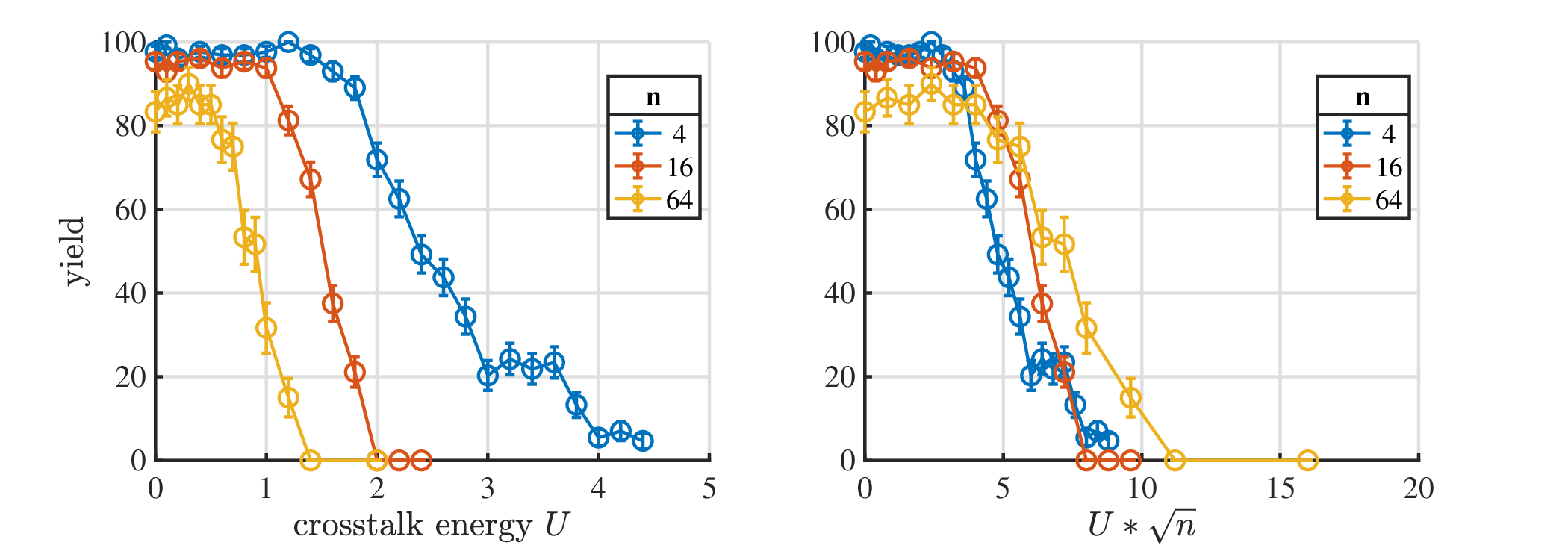}
\caption{Effect of crosstalk energy $c$ on the yield of perfect structures for different $n$. Right: raw data, with 1-standard deviation error bars. Left: same data with horizontal axis rescaled by $\sqrt{n}/\delta_1$. Parameters for simulations are shown in Table \ref{tbl:crosstalk}.}\label{fig:crosstalk}
\end{figure*}

We explored the effect of crosstalk by including a constant interaction energy $U$ between any pair of squares in contact that are not in their native positions. Figure \ref{fig:crosstalk} (left) shows the yield (perfect yield) as a function of $c$ for different $n$, and Figure \ref{fig:crosstalk} shows the yield as a function of $U/\delta_1 \sqrt{n}$, to explore how the threshholds $U_c$ change with $n$. For each $n$ we chose the value of $\delta_1$ which maximized the yield in Figure \ref{fig:n16}(c), and we chose times increasing by a factor of $8$ for each subsequent value of $n$, as for the simulations in Figure \ref{fig:n16}(c). 
The full set of parameters used is given in Table \ref{tbl:crosstalk}.

\begin{table}
\begin{tabular}{c | llll}
$n$ &  $\delta_1$ & time (MC sweeps) & $N_c$ & \# reps \\\hline
4 & 7.5 & $1.28\times 10^5$ & 128 & 1 \\
16 & 8.0 & $1.024\times 10^6$ & 128 & 1\\
64 & 8.5 & $8.192\times 10^6$ & 60 & 1\\
\end{tabular}
\caption{Parameters used for the crosstalk simulations in Figure \ref{fig:crosstalk}. All simulations used volume fraction $\phi=0.05$.}\label{tbl:crosstalk}
\end{table}


 \section{Testing different models of diffusion}\label{sec:diffusion}

 We explored different models for how the diffusivity of clusters depends on their size. 
 We start by calculating the ``hydrodynamic radius'' as 
 \[
 r_{\rm hydr} = r_0 + \langle |(r_i-r_c)\times \hat n|^2\rangle^{1/2}
 \] 
 where $r_0 = 1$ is the radius of a monomer, $\langle \cdot \rangle = \frac{1}{n_C}\sum_{i=1}^{n_C}$ is the average over the $n_C$ particles in a cluster, $r_i$ is the position of the $i$th particle in a cluster, and $r_c = \langle r_i\rangle$ is the center of mass of the cluster. 

 Our simulations reported in the main text use, as model for 3d Stokes' drag. The diffusion coefficient is 
 \[
 D = \frac{r_0}{r_{\rm hydr}}. 
 \]
 This model damps the diffusion in a way that is roughly inversely proportional to the cluster's transverse dimensions, so we call it diffusion with power $\alpha=1$. 

 In this section we briefly explore other models for diffusion. We decided to modify the power of the hydrodynamic radius, so that 
 \begin{equation}\label{Dalph}
 D = \frac{r_0}{( r_{\rm hydr})^\alpha}. 
 \end{equation}
 The models we consider are 
 \begin{enumerate}
 \item $\alpha = 0$. This sets $D=1$ no matter what size the cluster is. All clusters diffuse at the same rate. 
 \item $\alpha = 2$. Then $D = 1/r_{\rm hydr}^2$. Larger clusters diffuse significantly more slowly than smaller clusters. 
 \item $\alpha = -1$. Then $D \propto r_{\rm hydr}$. Since the maximum possible value of $D$ is 1, we divide by the largest possible value of $r_{\rm hydr}$, to obtain 
 \[
 D = \frac{r_{\rm hydr}}{r_{\rm max}}, \qquad r_{\rm max} = 1+\sqrt{n_0},
 \]
 where $n_0$ is the largest possible cluster we are forming (e.g. 4, 16, 64, etc). 
 In this model, large clusters diffuse significantly \emph{faster} than smaller clusters. 
 \end{enumerate}

\begin{figure}
 \includegraphics[width=0.5\textwidth]{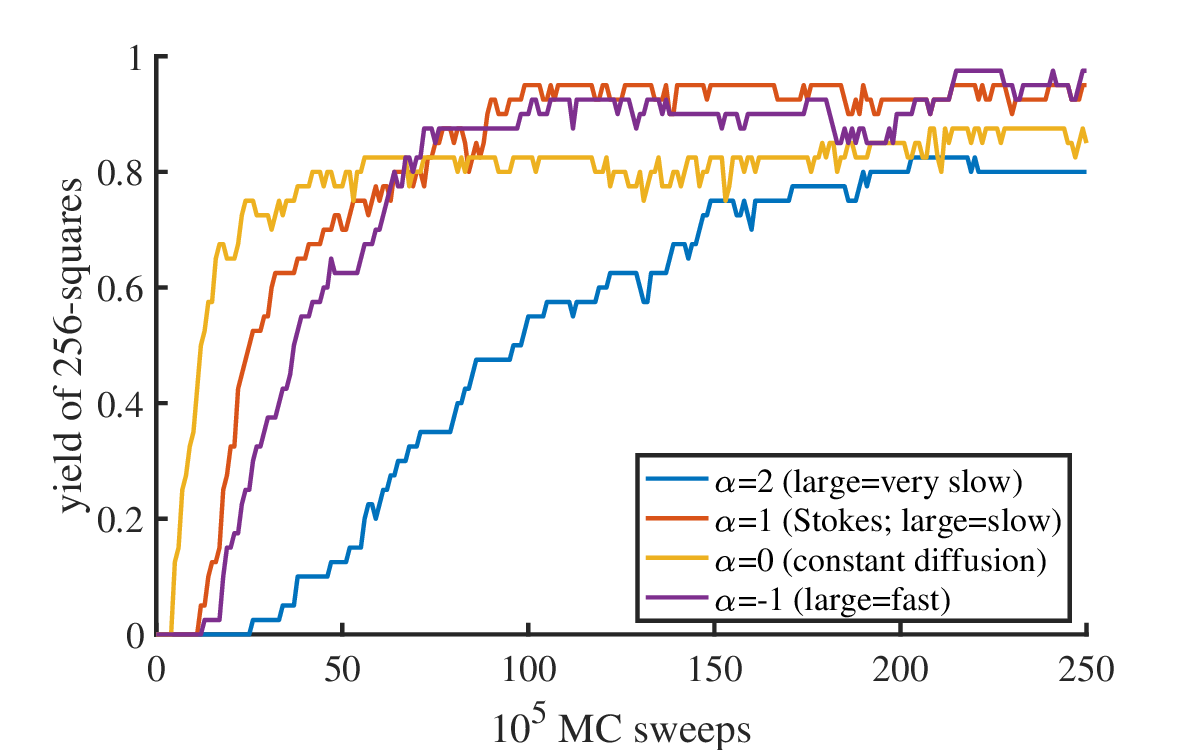}
 \caption{Yield of 256-squares as a function of time, using different diffusion models with different values of $\alpha$ in \eqref{Dalph}. All simulations considered 4 copies of a 256-square and were repeated 10 times each, with $\delta_1 = 8.75$ and hierarchical interactions.}\label{fig:diffusion}
 \end{figure}

 Figure \ref{fig:diffusion} shows the yield of 256-squares using different diffusion models. The models vary in their overall timescales, notably with the model with $\alpha=2$ assembling more slowly, presumably because the overall average diffusion coefficient is slower. However, the differences in the yields at some finite time is not dramatic, showing that the details of the diffusion are not important for hierarchical assembly to be successful. 

\end{document}